\documentclass[9pt,twocolumn,twoside]{osajnl}
\journal{ol} 
\setboolean{shortarticle}{false}
\usepackage{xfrac}
\usepackage{hyperref}
\usepackage{balance}
\usepackage{lineno}
\title{Low-coherence off-axis digital holographic microscopy}
\author[1,*]{Stephane Perrin}
\author[1,2]{Jonas K\"uhn}
\author[1,3]{Christian Depeursinge}

\affil[1]{Advanced Photonics Laboratory, Ecole Polytechnique Fédérale de Lausanne, CH-1015 Lausanne, Switzerland}
\affil[2]{Department of  Astronomy, University of Genève, CH-1211 Genève, Switzerland}
\affil[3]{Laboratory for Cellular Imaging and Energetics, King Abdullah University of Science and Technology, Thuwal, Kingdom of Saudi Arabia}

\affil[*]{Corresponding author: stephane.perrin@epfl.ch}

\begin{abstract}
Usually, off-axis digital holographic microscopy requires a coherent light source in order to record a full-field hologram. Nevertheless, a LASER-based illumination leads to a non-negligible coherent noise, decreasing then the imaging quality. We hereby report a simple method to reduce the coherent noise contribution using a low-temporal-coherence illumination while maintaining a large interference area. A diffraction grating is hence introduced in the reference arm of the interferometer, allowing the coherence plane of the reference beam to be tilted following angular dispersion. The phase planes of the reference beam and the object beam appears to be coplanar. The principle and performance of low-coherence off-axis digital holographic microscopy are demonstrated. The three-dimensional reconstruction of a biological sample is performed.
\end{abstract}

\setboolean{displaycopyright}{true}

\begin{document}

\maketitle
Since the demonstration of recording a modulated interference signal using a vidicon detector in 1967 by W. Goodman \cite{Goodman1967}, digital holography has been improved through the enhancement of sensors and computing power \cite{Schnars94}. At the end of the nineties, digital holography was combined to optical microscopy, leading to the development of a non-invasive three-dimensional imaging technique, called digital holographic microscopy (DHM) \cite{Depeursinge99,Ferraro03,Kim11,Osten2014}. DHM is nowadays widely employed for the visualization of biological samples \cite{Marquet05,Kemper06,Bally2008,Molder08,Yi17,Yi16} as well as for the characterization of micro-optical elements \cite{Jueptner01,Charriere06,Ferraro06,Weijuan10,Kozacki11}. DHM provides an interferometric nanometeric accuracy \cite{Kuhn08} and a lateral resolution limited by the diffraction of light \cite{Goodman04}. Recently, \textit{super-resolution} methods has even been implemented in order to perform a sub-wavelength transversal resolution \cite{Gao13,Cotte13,Aakhte17,Lin18}. Nevertheless, DHM is subject to coherent noise due to the use of spatially and temporally coherent light sources \cite{Chavel80}. As a matter of fact, the reflections of coherent beams by multiple optical interfaces generate undesirable interference signals, contaminating the hologram and thus introducing unwanted artefacts in the reconstructed image.
\par In the last decade, many approaches have been developed in order to remove the coherent noise such as recording wavelength-dependent holograms \cite{Nomura08}, displacing or rotating the sample \cite{Pan11,Zhong17}, implementing a syntetic illumination aperture \cite{Feng09} and processing the images \cite{Liu16}. Ref.~\cite{Bianco18} reports these best-performing noise reduction approaches. However, these techniques present some drawbacks, \emph{i.e.} reducing the resolution of the imaging system, decreasing the contrast of the interference pattern or necessitating multiple acquisitions. The coherent noise only occurs when the distance between two nearby interfaces is smaller than the coherence length of the light source. Thus, decreasing the spatial or the temporal coherence of light has also been suggested \cite{Schedin01,Dubois06,Kemper10}. In addition, using an incoherent illumination tends to increase the lateral resolution of an imaging system \cite{Goodman04}. Nevertheless, this implies narrowing the width of the interference pattern in off-axis DHM configuration. At the begining of 1990's, it has been shown that angular dispersion from a diffraction grating (or a prism) provides a tilt of the pulse-front while not affecting the group velocity \cite{Martinez1986,Szabo1993,Hebling1996,Maznev98}. Indeed, a diffraction grating not only separates an incident beam into several beams at different angles, but also tilts the pulse plane of a broadband incident beam \cite{Torres10}. In other words, the pulse front of a beam from a low-coherent light source appears no longer co-planar to the direction of the beam propagation after passing through a diffraction element, but inclined with respect to the diffraction order angle. This phenomenon gave the opportunity to make the coherence planes of two angled incoherent beams parallel themselves and, applied to holography, made it possible the full-field single-shot hologram acquisition using two prisms \cite{Ansari01}, a diffraction grating \cite{Yaqoob11} or multiple gratings combined with a dual-wavelengths illumination \cite{Monemhaghdoust11,Monemhaghdoust13}.
\par This letter presents an easy-to-implement method enabling to enhance the experimental conditions in transmissive off-axis DHM. Indeed, the interference system requires a low-coherence illumination and a standard diffraction grating in the reference arm for making coplanar the beam coherence planes. The concept of low-coherence off-axis DHM is exposed and the increase in interference pattern width is demonstrated. Furthermore, performance of the imaging technique is estimated. Then, the volume of a neuron is reconstructed.
\par
\begin{figure}[!h]
\centering
\includegraphics[width=\linewidth]{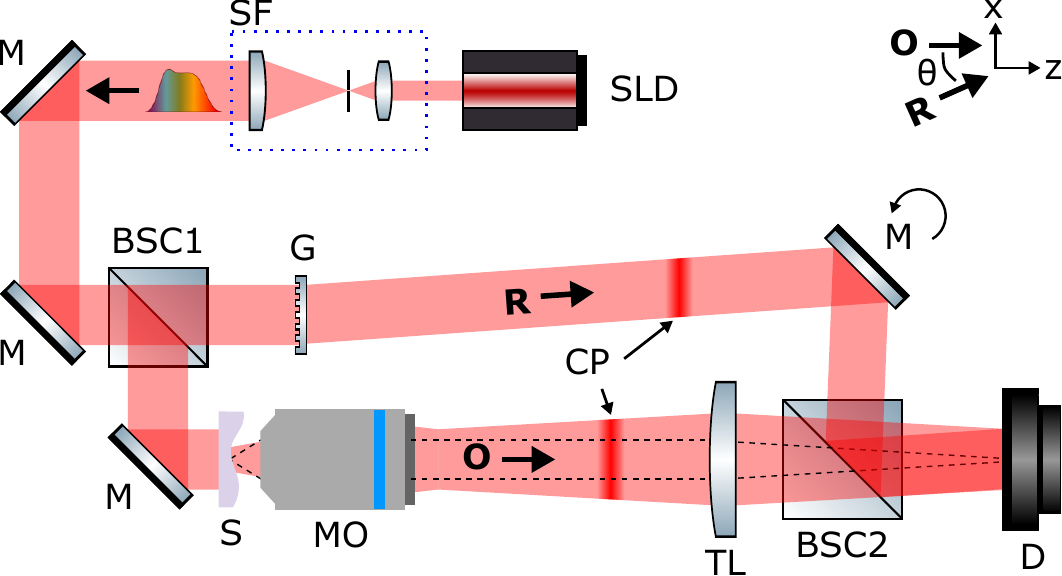}
\caption{Experimental configuration of the low-coherence off-axis DHM. SLD, Super-luminescent diode; SF, spatial filter; M, mirror; BSC, beam-splitter cube; G, diffraction grating; CP, coherence plane; S, sample; MO, microscope objective; TL, tube lens; D, detector.}
\label{fig:fig1}
\end{figure}

 The layout of the off-axis DHM setup is shown in Fig.~\ref{fig:fig1} and consists of a super-luminescent diode having a central wavelength $\lambda_{0}$ of 680~nm and a bandwidth $\Delta\lambda$ of 8~nm (SLD-26-HP, Superlum Diode). The incident beam is firstly spatially filtered before being divided in two by a beam-splitter cube. The reference beam \textbf{R} of the interferometer is transmitted and passes through a diffraction grating (46-067, Edmund Optics) having a groove density of 70~lines/mm. According to angular dispersion, the coherence plane is tilted by an angle $\Phi(\lambda_{0})$ of 2.7° with respect to the propagating vector \cite{Szabo1993}. Only the $+1$-order diffracted beam from the grating is required in the imaging system. Other order beams are removed using an obstruction which is not represented in Fig.~\ref{fig:fig1}. 
The beam reflected by the cube, \textit{i.e.} the object beam \textbf{O}, is oriented by a mirror in order to illuminate the specimen to be measured. The microscope objective ($\times$5, NA~=~0.17) collects the beam scattered by the sample and the tube lens forms the image on the CCD camera (TXG50, Baumer). The second beam-splitter cube allows to surimpose the \textbf{O} beam with the \textbf{R} beam in the off-axis mode. Therefore, the mirror in the reference arm is able to adjust the incident angle $\theta$ between the two beams. The interferogram distribution is then recorded by the camera and, finally, an algorithm processes the full-field hologram and performs a numerical wavefront propagation using 2D Fast Fourier Transform (FFT) operators \cite{Depeursinge99}. This numerical operation makes allows to retrieve the phase distribution of the sample.
\par Without the diffraction element in the reference arm, the width of interference area $L$ is limited by the coherence length of the light source $l_{c}$ and the angle $\theta$ between the two beams ($L=l_{c}/\sin\left(\theta\right)$). Assuming a Gaussian-spectrum light illumination, the number of interference fringes $N$, having an interference contrast superior to 50\%, can be expressed as \cite{Saleh07}:
\begin{equation}
N = \frac{2 \ln(2)}{\pi}~\frac{\lambda_{0}}{\Delta \lambda}
\label{Eq1}
\end{equation}
\begin{figure}[!b]
\centering
\includegraphics[width=\linewidth]{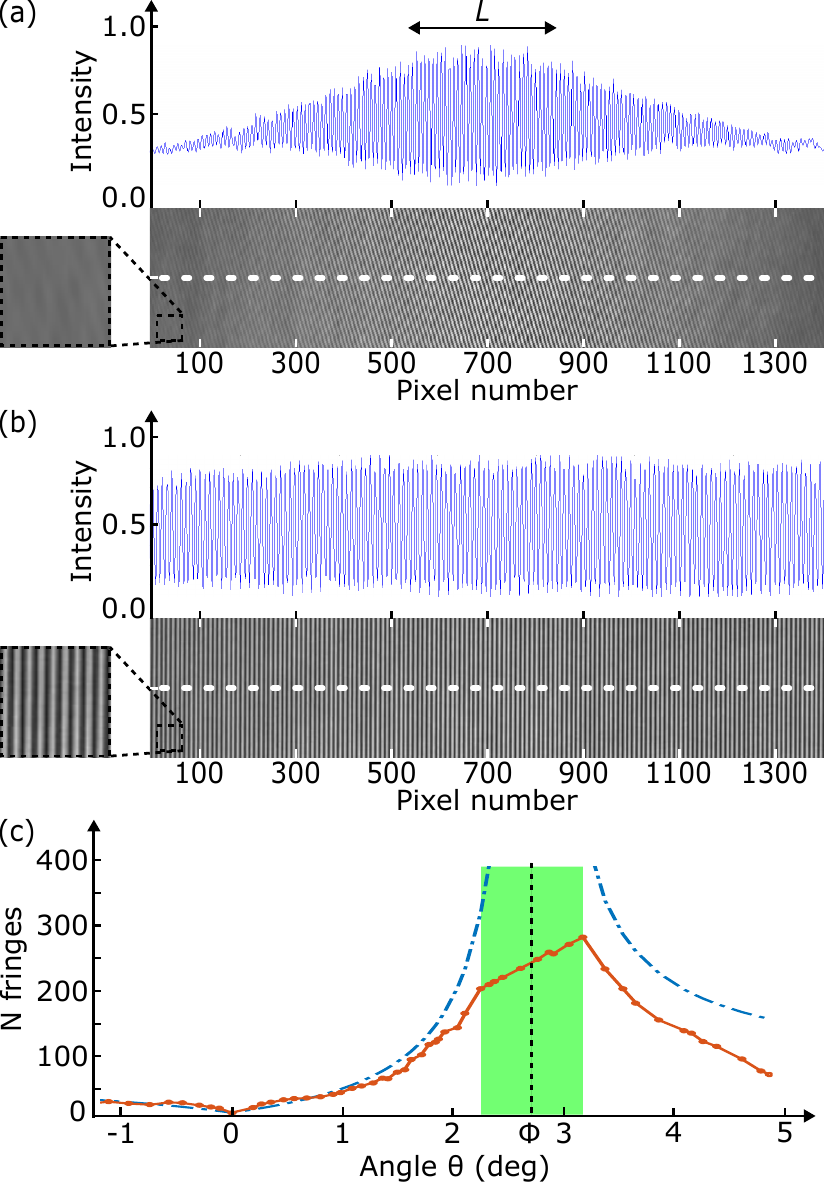}
\caption{Increase in the interference area size. (\textbf{a}) Overlap of two incoherent beams in conventional off-axis DHM ($\theta$~=~2.7°). The interference width $L$ equals 533~µm. (\textbf{b}) Overlap of the two incoherent beams by placing a diffraction grating in the reference arm. The camera records an interference pattern over the whole sensor surface. The first-order diffraction angle of the grating $\Phi(\lambda_{0})$ equals the angle $\theta$. The normalized intensity profiles are plotted according to the white doted lines. (\textbf{c}) Evolution of both the calculated (in blue) and the measured (in red) number of fringes $N$ as a function of the angle $\theta$. Within the green area, $L$ is wider than the sensor size.}
\label{fig:fig2}
\end{figure}
In this case, the camera records only 38 interference fringes, providing thus fringe-less observation regions as shown in Fig.~\ref{fig:fig2}(a). The diffraction grating is used to tilt the Poynting vector with respect to the propagation direction. And, when $\theta$ equals $\Phi(\lambda_{0})$, the coherence planes are collinear and an uniform interference contrast is thus recorded over the entire sensor area. Figure~\ref{fig:fig2}(b) shows the resulting 2D interference pattern, performing a mean interference contrast of 80\%. The number of fringe is now defined as a function of the angles $\theta$ and $\Phi$.
\begin{equation}
N = \frac{2 \ln(2)}{\pi}~\frac{\lambda_{0}}{\Delta \lambda}~\frac{\sin\left(\theta\right)}{\sin\left(\Phi(\lambda)~-~\theta\right)}
\label{Eq2}
\end{equation}
Figure \ref{fig:fig2}(c) shows the evolution of the number of fringes $N$ from Eq.~\ref{Eq2} (blue dashed plot) as a function of the incident angle $\theta$. When $\theta$ equals $\Phi(\lambda_{0})$, an infinity number of fringes should be recorded by the camera. However, in experiment (red solid plot), the 5-mm-size sensor limits the lateral field of view, yielding a linear evolution of $N$ between 2.2° and 3.2° (green area). For the measurements, the angle $\theta$ was implemented by rotating the mirror in the reference arm around the optical axis and the DHM system was free of object. Note that the incident angle $\theta$ depends also on the wavelength of light due to the bandwidth of the light source. This involves adjusting the grating-camera distance in order to record all the wavelength-dependent diffracted beams, \emph{i.e.} the contributions from each spectral component. Not considering this chromatic effect could lead to a fringe wash-out and further a loss of the interference contrast. In addition, at the ideal angular position, the angle $\theta(\lambda)$ equals $\Phi(\lambda)$ regardless the wavelength.
\par
\begin{figure}[!h]
\centering
\includegraphics[width=\linewidth]{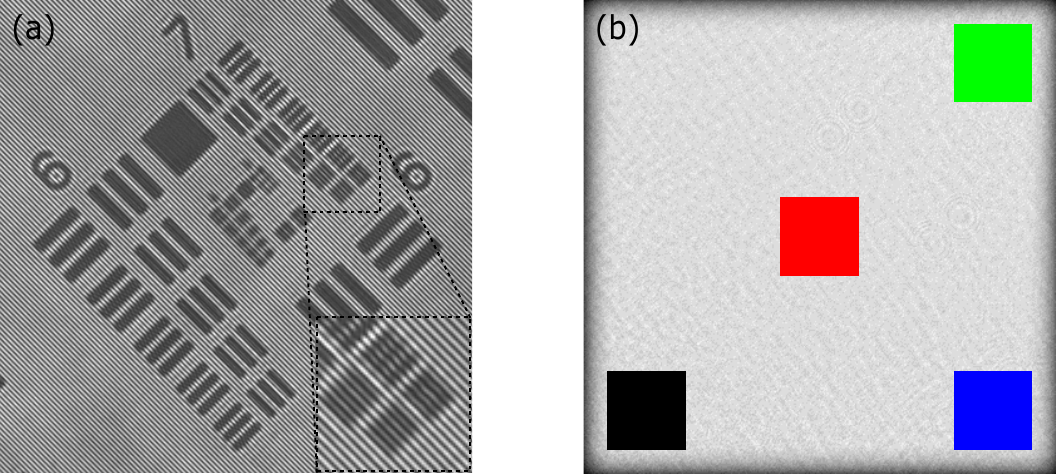}
\caption{Performance of low-coherence off-axis DHM system. (\textbf{a}) Hologram of a 1951 USAF target. The DHM system can resolve Element 6, Group 7. (\textbf{b}) Measurement of the spatial phase deviation using a flat mirror. The RMS of the phase deviation is 32~mrad within blue area, 30~mrad within black area and 27~mrad within green and red areas. Each square area consists of 100~$\times$~100 pixels. $\lambda_{0}$~=~680~nm, NA~=~0.17.}
\label{fig:fig3}
\end{figure}

 Performance of the low-coherence interference imaging system was evaluated. Figure~\ref{fig:fig3}(a) shows the hologram of a 1951 USAF target, allowing the lateral resolution to be determined. It results in 2.19~µm of resolution limit (Element 6, Group 7). This allows to highlight that the diffraction grating has not impact on the imaging quality and the DHM system is thus assumed diffraction limited (the cut-off frequency of the optical transfer function of an aberration-free imaging system equals 2.00~µm \cite{Goodman04}). Note that the resolving power of the imaging system would be around 4~µm using a 680-nm-wavelength monochromatic light source. 
\begin{figure}[!b]
\centering
\includegraphics[width=\linewidth]{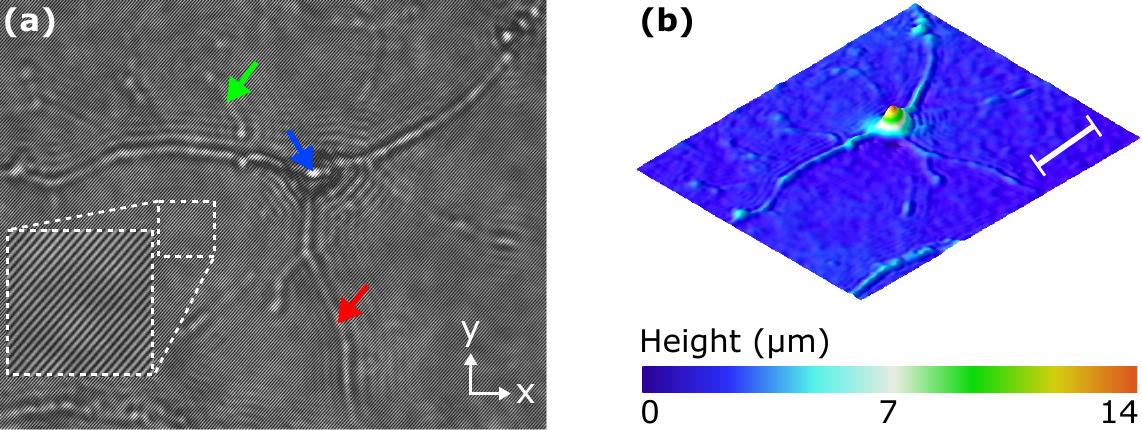}
\caption{3D reconstruction of a mouse's neuron using low-coherence off-axis DHM. (\textbf{a}) Width-field hologram. (\textbf{b}) reconstruction of the morphology. The soma, the dendrites and the axons of the neuron are recognized with the blue, the green and the red arrows, respectively. $\lambda_{0}$~=~680~nm, NA~=~0.17. White scale bar represents 100 µm.}
\label{fig:fig4}
\end{figure}
Furthermore, a quantitative analysis of the phase deviation has been measured (Fig.~\ref{fig:fig3}(b)) through the reconstruction of a plane mirror. The root mean square (RMS) of the phase noise equals 30~mrad, \textit{i.e.} 3.2~nm of height deviation, over the entire field of view. Usually, DHM provides an axial sensibility of around 10~nm \cite{Depeursinge99,Monemhaghdoust13,Tobin08}
\par Finally, a biological element has been introduced in the low-coherence off-axis DHM system for tracking the shape and the cellular behaviour. Figure~\ref{fig:fig4}(a) shows the width-field hologram of the fixed mouse neuron recorded by the camera. The quantitative phase distribution was then calculated, enabling the morphology of the brain cell to be reconstructed (Fig.~\ref{fig:fig4}(b)). Assuming a mean refractive index of 1.375 along the propagation axis inside the mouse brain cell \cite{Marquet05,Marquet13}, the height of at the center of the neuronal cell body (blue arrow in Fig. \ref{fig:fig4}(a)), i.e. the soma, equals 12.9~µm. And, the height of the dendrites (green arrow in Fig. \ref{fig:fig4}(a)) is around 3.8~µm. 
\par This work presents the development of a low-coherence off-axis digital holographic microscope. Illuminated by a broadband light source, the parasitic coherent noise contributions which originate from multiple reflections between the optical component interfaces, are reduced. Despite, this type of illumination narrows the hologram width in off-axis configuration, a diffraction grating has been introduced in the reference arm of the transmissive-configuration interferometer in order to make parallel the reference and the object wave-packets. This method allows the hologram to cover the camera and, further, the standard deviation of the phase reconstruction to be lower than 30~mrad. The imaging technique has been applied for the reconstruction of a neuron cell. It can be mentioned that this method can be implemented in a reflective configuration.
\begin{backmatter}
\bmsection{Funding} This research was funded by the Swiss National Science Foundation (SNSF) and has been supported by King Abdullah University of Science and Technology. This work has been made in collaboration with the Microvision Microdiagnostic Group of the Ecole Polytechnique Federale de Lausanne and Lyncee Tec SA.
\bmsection{Disclosures} The authors declare no conflicts of interest.
\end{backmatter}

\end{document}